\documentclass[english,two column]{revtex4-1}
\usepackage[T1]{fontenc}
\usepackage[latin9]{inputenc}
\usepackage{color}
\usepackage{amsmath}
\usepackage{amssymb}
\usepackage{graphicx}
\usepackage{babel}
\begin{document}
\title{Dynamic structure factors of a strongly interacting Fermi superfluid
near an orbital Feshbach resonance across the phase transition from
BCS to Sarma superfluid}
\author{Peng Zou$^{1}$}
\email{phy.zoupeng@gmail.com}

\author{Huaisong Zhao$^{1}$, Lianyi He$^{2}$, Xia-Ji Liu$^{3}$, and Hui
Hu$^{3}$ }
\affiliation{$^{1}$College of Physics, Qingdao University, Qingdao 266071, China}
\affiliation{$^{2}$Department of Physics, Tsinghua University, Beijing 100084,
China}
\affiliation{$^{3}$Centre for Quantum Technology Theory, Swinburne University
of Technology, Melbourne 3122, Australia}
\begin{abstract}
We theoretically investigate dynamic structure factors of a strongly
interacting Fermi superfluid near an orbital Feshbach resonance with
random phase approximation approach at zero temperature, and find
their dynamical characters across the phase transition from a balanced
conventional Bardeen-Cooper-Schrieffer superfluid to a polarized Sarma
superfluid by continuously varying the chemical potential difference
of two spin components. In a BEC-like regime of the Fermi superfluid,
dynamic structure factors can help distinguish the in-phase ground
state from the out-of-phase metastable state by the relative location
of molecular excitation and Leggett mode, or the minimum energy to
break a Cooper pair. In the phase transition from BCS to Sarma superfluid,
we find the dynamic structure factor of Sarma superfluid has its own
specific gapless excitation at a small transferred momentum where
the collective phonon excitation acquires a finite width, and also
a relatively strong atomic excitation at a large transferred momentum,
because of the existence of unpaired Fermi atoms. Our results can
be used to differentiate Sarma superfluid from BCS superfluid.
\end{abstract}
\maketitle

\section{Introduction}

Interacting quantum many-body system is always a central research
field in physics, and contains many interesting but also challenging
problems. As a many-body physical quantity, the density-density dynamic
structure factor, i.e., the Fourier transformation of density-density
correlation function, contains rich information about the dynamical
excitation of the system \cite{pitaevskiibook}. At a low transferred
momentum regime, one can observe many kinds of collective excitations,
while molecular and atomic excitation can be seen at a large transferred
momentum. Besides the spectrum of the system, dynamic structure factors
also display the weight distribution of both Fermi atoms and Cooper-pairs
according to the width and strength of excitation peak. From both
the spectrum and weight distribution we can understand rich many-body
properties, like the pair condensation and transition temperature
\cite{lingham2014}, Tan's contact parameter \cite{kuhnle2011,hoinka2013},
\textit{etc},. Generally the experimental measurement of many-body
physical quantities is a quite difficult task, luckily a two-photon
Bragg scattering technique can be able to measure dynamic structure
factors at an almost full regime of transferred momentum \cite{veeravalli2008,hoinka2017,hoinka2012,brunello2001}
.

The accumulation of interacting many particles can form lots of quantum
matter states, \textit{i.e}. Fulde-Ferrell-Larkin-Ovchinnikov (FFLO)
phases \cite{fulde1964,larkin1964}, Sarma superfluid \cite{sarma1963},
topological superfluid \cite{zhang2008} \textit{etc},. The search
and distinguishment of new matter states in different materials or
microscopic particles system are interesting and important jobs. The
BCS-type Fermi superfluid is a conventional superfluid state, where
two Fermi atoms with the same amplitude of momentum but unlike direction
and spin can form a Cooper pair with zero momentum of center of mass.
It is usually the ground state of many superconductors or Fermi superfluid.
A Sarma superfluid is a possible candidate ground state for a spin-polarized
Fermi superfluid. Having gapless fermionic excitations, this state
can be viewed as a phase separation state in momentum space: some
fermions pair and form a superfluid, while other unpaired atoms just
occupy in certain regions of momentum space bounded by gapless Fermi
surfaces. 

In the context of two-component spin-1/2 Fermi atomic gas at the famous
BEC-BCS crossover, Sarma superfluid can only be stable on the BEC
side of the crossover. It may also be stabilized by considering a
multiband structure \cite{he2009}, which experimentally can be realized
with alkali-earth-metal atoms (such as Sr) or alkali-earth-metal-like
atoms (i.e. Yb), where there are two electrons outside the atoms.
In this case the usual Magnetic Feshbach resonance fails to tune the
effective interacting strength since their zero magnetic moment difference,
but a pioneering work by R. Zhang {\it{et al}} theoretical proposed
an alternative mechanism of orbital Feshbach resonance (OFR) for $^{173}$Yb
\cite{zhang2015}. In Fermi gases of $^{173}$Yb, the long-lived metastable
electric orbital state $^{3}P_{0}$ (denoted as $\left|e\sigma\right\rangle $
where $\sigma=\uparrow,\downarrow$ stands for the two internal nuclear
spin states) can be chosen, together with the ground orbital state
$^{1}S_{0}$ ($\left|g\sigma\right\rangle $) . In fact this forms
a four-component Fermi system, where a pair of atoms can be described
by using the singlet ($-$) and triplet ($+$) basis in the absence
of external Zeeman field,

\begin{equation}
\left|\pm\right\rangle =\frac{1}{2}\left(\left|ge\right\rangle \pm\left|eg\right\rangle \right)\otimes\left(\left|\uparrow\downarrow\right\rangle \mp\left|\downarrow\uparrow\right\rangle \right),
\end{equation}
or by a usual two-channel basis in the presence of the Zeeman field

\begin{equation}
\left|o\right\rangle =\frac{1}{\sqrt{2}}\left(\left|-\right\rangle +\left|+\right\rangle \right)=\frac{1}{\sqrt{2}}\left(\left|g\uparrow,e\downarrow\right\rangle -\left|e\downarrow,g\uparrow\right\rangle \right),
\end{equation}

\begin{equation}
\left|c\right\rangle =\frac{1}{\sqrt{2}}\left(\left|-\right\rangle -\left|+\right\rangle \right)=\frac{1}{\sqrt{2}}\left(\left|g\downarrow,e\uparrow\right\rangle -\left|e\uparrow,g\downarrow\right\rangle \right),
\end{equation}
where $o$ and $c$ are short for open and close channel, respectively.
The interparticle interaction is described by a $s$-wave scattering
length, namely the singlet scattering length $a_{-}$ and the triplet
one $a_{+}$. Because of a shallow bound state (related to a large
$a_{+}\simeq1900a_{0}$, where $a_{0}$ is the Bohr radius) caused
by the interorbital (nuclear) spin-exchange interaction, the small
difference in the nuclear Land\'e factor between different orbital
states allows the tunability of scattering length in the open channel
through an external magnetic field\cite{zhang2015}. The existence
of the predicted OFR has been confirmed soon experimentally \cite{pagano2015,hofer2015}
. More fascinatingly, OFR is a narrow resonance due to the significant
closed-channel fraction \cite{xu2016}, and a two-channel mode \cite{he2015}
is really necessary to provide a right description. In the two-channel
system, the exchange interaction between two channels varies significantly
the stability mechanism of the pairing state, and makes Sarma state,
which is always unstable in single-channel system, can be a stable
ground state \cite{he2009} once a nonzero interband exchange interaction
and a large asymmetry between the two bands are existent. In our recent
work, a superfluid phase transition from a balanced conventional Bardeen-Cooper-Schrieffer
superfluid to a polarized Sarma superfluid in a strongly interacting
Fermi superfluid near an orbital Feshbach resonance has been reported
\cite{zou2018}. 

It is of great interest to explore the many-body physical eigenstates
of OFR, and investigate their rich dynamics. The existence of two
order parameters in OFR has opened the possibility of observing the
long-sought {\it{massive}} Leggett mode resulted from the fluctuation
of the relative phase of the two order parameters, which can be observed
in both BCS and Sarma superfluid \cite{leggett1966,blumberg2007,lin2012,bittner2015,klimin2019}.
So it is interesting to study all dynamics of OFR , and find their
specific characters in different matter states, from which to research
possibility to distinguish different many-body eigenstates, or different
ground states of OFR by an experimental observable many-body physical
quantity. Following this idea, we would like to investigate the dynamic
structure factor across the phase transition process from BCS superfluid
to Sarma superfluid in OFR, and find their own specific dynamical
characters which may service as fingerprints of different matter states.

This paper is organized as follows. In the next section, we will use
the language of Green's function to introduce the microscopic model
of a three-dimensional Fermi superfluid near OFR with both spin-balanced
and spin-polarized population , outline the mean-field approximation
and how to calculate response function with random phase approximation,
and give the results of dynamic structure factor of BCS superfluid
in Sec. III, and the results of Sarma superfluid in Sec. IV. In Sec.
V and VI, we will give our conclusions and acknowledgement, respectively.

\section{Mean-field theory}

\subsection{Model Hamiltonian}

The spin-population imbalanced Fermi gases near orbital Feshbach resonance
can be described by a two-channel Hamiltonian 

\begin{equation}
\mathcal{H}=\int d{\rm \textit{\textbf{r}}}\left[\underset{ni}{\sum}\psi_{ni}^{\dagger}\mathcal{H}_{ni}^{0}\psi_{ni}+\underset{nm}{\sum}V_{nm}\varphi_{n}^{\dagger}\varphi_{m}\right]
\end{equation}
where $\mathcal{H}_{ni}^{0}=-\hbar^{2}\nabla^{2}/2M-\mu_{ni}$ is
the single particle Hamiltonian of spin $i=1,2$ Fermi atoms with
mass $M$ and chemical potential $\mu_{ni}$, the subscript $n=o,c$
denotes the open or close channel, and $\psi_{ni}\left(\textit{\textbf{r}}\right)$
and $\psi_{ni}^{\dagger}\left(\textit{\textbf{r}}\right)$ are the
corresponding annihilation and creation field operator of Fermi atoms,
respectively. $\varphi_{n}\equiv\psi_{n2}\psi_{n1}$ is an anomalous
density operator. In each channel we assume spin-up chemical potential
$\mu_{n1}=\mu_{n}+\delta\mu_{n}$ and spin-down one $\mu_{n2}=\mu_{n}-\delta\mu_{n}$,
where $\delta\mu$ is the chemical potential difference of different
spin around an average value $\mu_{n}$. In the presence of a Zeeman
field, a pair of atoms in the open and close channel has different
Zeeman energy with a difference $\delta\left(B\right)=\left(g_{g}m_{2}+g_{e}m_{1}\right)\mu_{B}B-\left(g_{g}m_{1}+g_{e}m_{2}\right)\mu_{B}B$,
arising from their difference in magnetic momentum. So we may define
the effective chemical potentials of the open and close channel as
$\mu_{o}=\mu$ and $\mu_{c}=\mu-\delta\left(B\right)/2$. The interaction
potentials between atomic pairs can be well approximated by $s$-wave
contact pseudopotentials 

\begin{equation}
V_{\pm}\left(r\right)\simeq\frac{4\pi\hbar^{2}a_{\pm}}{M}\delta\left(r\right).
\end{equation}
 When using the external Zeeman field as a control knob, it will be
convenient to use the two-channel language to describe contact interaction
potentials, which become

\begin{equation}
V_{oo}\left(r\right)=V_{cc}\left(r\right)=\frac{V_{-}+V_{+}}{2}=\frac{4\pi\hbar^{2}a_{0}}{M}\delta\left(r\right),
\end{equation}

\begin{equation}
V_{oc}\left(r\right)=V_{co}\left(r\right)=\frac{V_{-}-V_{+}}{2}=\frac{4\pi\hbar^{2}a_{1}}{M}\delta\left(r\right).
\end{equation}
The two scattering lengths $a_{0}$ and $a_{1}$ are given by $a_{0}=\left(a_{-}+a_{+}\right)/2$
and $a_{1}=\left(a_{-}-a_{+}\right)/2$, and the bare interaction
strengths $V_{nm}$ ($n,m=o,c$) need to be regularized with the two
scattering lengths $a_{0}$ and $a_{1}$ by the standard renormalization
procedure

\begin{equation}
\left[\begin{array}{cc}
V_{oo} & V_{oc}\\
V_{co} & V_{cc}
\end{array}\right]^{-1}=\frac{M}{4\pi\hbar^{2}}\left[\begin{array}{cc}
a_{0} & a_{1}\\
a_{1} & a_{0}
\end{array}\right]^{-1}-\sum_{k}\frac{M}{\hbar^{2}k^{2}}.
\end{equation}
Following the standard mean-field theoretical treatment, we may introduce
two order parameters in open and close channel

\begin{equation}
\left[\begin{array}{c}
\triangle_{o}\left(r\right)\\
\triangle_{c}\left(r\right)
\end{array}\right]\equiv\left[\begin{array}{cc}
V_{oo} & V_{oc}\\
V_{co} & V_{cc}
\end{array}\right]\left[\begin{array}{c}
\left\langle \varphi_{o}\left(r\right)\right\rangle \\
\left\langle \varphi_{c}\left(r\right)\right\rangle 
\end{array}\right].\label{eq:gap}
\end{equation}
After a Fourier transformation to field operator $\psi_{ni}=\sum_{k}c_{ni,k}\phi_{k}$
($\phi_{k}$ is the plane wave function), we obtain the mean-field
Hamiltonian in the momentum representation

\begin{equation}
\begin{array}{ll}
\mathcal{H}_{{\rm mf}} & =\underset{nk}{\sum}\left[\left(\xi_{nk}-\delta\mu\right)c_{n1,k}^{\dagger}c_{n1,k}+\left(\xi_{nk}+\delta\mu\right)c_{n2,k}^{\dagger}c_{n2,k}\right]\\
 & +\underset{nk}{\sum}\left(\Delta_{n}c_{n1,k}^{\dagger}c_{n2,-k}^{\dagger}+\Delta_{n}^{*}c_{n2,-k}c_{n1,k}\right)
\end{array}
\end{equation}
where $\xi_{nk}=\hbar^{2}k^{2}/2M-\mu_{n}$, and order parameter satisfies
$\Delta_{n}=\sum_{m}V_{nm}\sum_{k}\left\langle c_{m2,-k}c_{m1,k}\right\rangle $.
Since we just discuss the eigenstate, two order parameters should
be both real number and satisfy $\Delta_{n}\equiv\Delta_{n}^{*}$.

\begin{figure}
\includegraphics[scale=0.36]{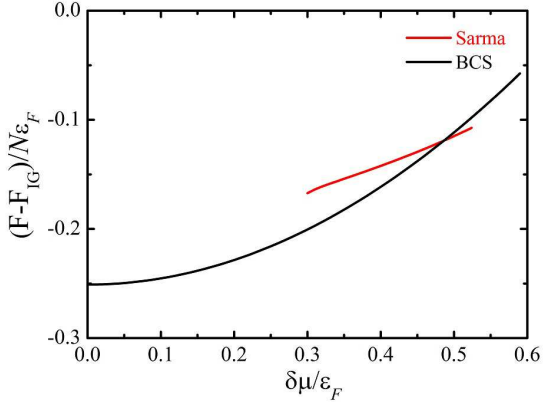}

\caption{\label{Fig1_phase} The free energy of both spin-population balanced
BCS state (black line) and spin-polarized Sarma state (red line) at
scattering length $\ensuremath{1/k_{F}a_{0}=-0.5}$, $\ensuremath{1/k_{F}a_{1}=-0.05}$
and Zeeman detuning $\delta(B)=0.4\varepsilon_{F}$. $F_{{\rm IG}}$
is the free energy of ideal Fermi gases with the same particle number.}
\end{figure}

With the language of Green's function, we start with equations of
motion of any operators $A$ and $B$ in the momentum-energy representation
$\omega\left\langle \left\langle A|B\right\rangle \right\rangle _{\omega}=\left\langle \left[A,B\right]_{+}\right\rangle +\left\langle \left\langle \left[A,\mathcal{H}_{{\rm mf}}\right]|B\right\rangle \right\rangle _{\omega}$,
and obtain three independent Green's function

\begin{equation}
\begin{array}{cll}
G_{1}^{\left(n\right)}\left(K\right) & \equiv\left\langle \left\langle c_{n1,k}|c_{n1,k}^{\dagger}\right\rangle \right\rangle _{\omega} & =\underset{l}{\sum}\frac{\frac{1}{2}\left(1+l\frac{\xi_{nk}}{E_{nk}}\right)}{i\omega-E_{nk}^{l}},\\
G_{2}^{\left(n\right)}\left(K\right) & \equiv\left\langle \left\langle c_{n2,k}|c_{n2,k}^{\dagger}\right\rangle \right\rangle _{\omega} & =\underset{l}{\sum}\frac{\frac{1}{2}\left(1-l\frac{\xi_{nk}}{E_{nk}}\right)}{i\omega+E_{nk}^{l}},\\
\Gamma^{\left(n\right)}\left(K\right) & \equiv\left\langle \left\langle c_{n1,k}|c_{n2,-k}\right\rangle \right\rangle _{\omega} & =\underset{l}{\sum}\frac{l\frac{\Delta_{n}}{2E_{nk}}}{i\omega-lE_{nk}^{l}},
\end{array}
\end{equation}
where $K=\left(k,i\omega\right)$ is a four-dimensional momentum,
and $\begin{array}{cc}
E_{nk}^{l}=-\delta\mu+lE_{nk}, & E_{nk}=\sqrt{\xi_{nk}^{2}+\Delta_{n}^{2}}\end{array},$ $l=\pm1$ stands for the particle or hole solution. These Green's
functions should be self-consistently solved with density equations

\begin{equation}
\rho_{o}=\sum_{k}\left\{ 1-\frac{\xi_{ok}}{E_{ok}}\left[1-f\left(E_{ok}^{+1}\right)-f\left(-E_{ok}^{-1}\right)\right]\right\} ,
\end{equation}

\begin{equation}
\rho_{c}=\sum_{k}\left\{ 1-\frac{\xi_{ck}}{E_{ck}}\left[1-f\left(E_{ck}^{+1}\right)-f\left(-E_{ck}^{-1}\right)\right]\right\} ,
\end{equation}
and order parameters equations

\begin{equation}
\frac{\lambda_{1}\Delta_{o}}{\Delta_{c}}+\lambda_{0}=\sum_{k}\left[\frac{M}{\hbar^{2}k^{2}}-\frac{1-f\left(E_{ck}^{+1}\right)-f\left(-E_{ck}^{-1}\right)}{2E_{ck}}\right]
\end{equation}

\begin{equation}
\frac{\lambda_{1}\Delta_{c}}{\Delta_{o}}+\lambda_{0}=\sum_{k}\left[\frac{M}{\hbar^{2}k^{2}}-\frac{1-f\left(E_{ok}^{+1}\right)-f\left(-E_{ok}^{-1}\right)}{2E_{ok}}\right]
\end{equation}
where 
\begin{equation}
\begin{array}{cc}
\lambda_{0}=+\frac{Ma_{0}}{4\pi\hbar^{2}\left(a_{0}^{2}-a_{1}^{2}\right)}, & \lambda_{1}=-\frac{Ma_{1}}{4\pi\hbar^{2}\left(a_{0}^{2}-a_{1}^{2}\right)}\end{array},
\end{equation}
and $f\left(x\right)=1/\left(e^{x/k_{B}T}+1\right)$ is the famous
Fermi-Dirac distribution function. The free energy of the system in
mean-field approximation reads

\begin{equation}
\begin{array}{cl}
F_{{\rm mf}}= & -\Delta^{\dagger}\left[\begin{array}{cc}
\lambda_{0} & \lambda_{1}\\
\lambda_{1} & \lambda_{0}
\end{array}\right]\Delta+\underset{nk}{\sum}\left(\xi_{nk}-E_{nk}+\frac{\Delta_{n}^{2}M}{\hbar^{2}k^{2}}\right)\\
 & -k_{B}T\underset{nkl}{\sum}{\rm ln}\left(1+e^{-E_{nk}^{l}/k_{B}T}\right)+\mu N
\end{array}
\end{equation}
where $\Delta\equiv\left[\begin{array}{cc}
\Delta_{o}, & \Delta_{c}\end{array}\right]^{T}$. As shown in Fig.\ref{Fig1_phase}, a phase transition from a spin-balanced
BCS superfluid to a spin-polarized Sarma superfluid is observed by
continuously increasing chemical potential difference $\delta\mu$
over a critical value $\delta\mu_{{\rm cr}}=0.486\varepsilon_{F}$. 

\subsection{Random phase approximation and response function}

The fluctuation term of interaction Hamiltonian plays a great role
in the calculation of dynamics of the system, random phase approximation
has been proved to be an effective way to collect the contribution
from the fluctuation term. To introduce this story about how to obtain
an accurate response function with random phase approximation approach,
let us begin with the linear response theory. 

In any superfluid state of OFR, there are usually four different densities
in both open and close channels. Besides the normal spin-up and down
densities ($\rho_{i}^{n}\equiv\left\langle \psi_{ni}^{\dagger}\psi_{ni}\right\rangle $,
$i=1,2$), the other two are anomalous density $\rho_{3}^{n}\equiv\left\langle \psi_{n2}\psi_{n1}\right\rangle $
and its complex conjugate $\rho_{4}^{n}\equiv\left\langle \psi_{n1}^{\dagger}\psi_{n2}^{\dagger}\right\rangle $.
When a weak external perturbation potential $V_{{\rm ext}}$ is imposed
to all densities in two channels with form $V_{{\rm ext}}=\left[\begin{array}{cc}
V^{o}, & V^{c}\end{array}\right]^{T}$ and $V^{n}=\left[\begin{array}{cccc}
V_{1}^{n}, & V_{2}^{n}, & V_{3}^{n}, & V_{4}^{n}\end{array}\right]^{T}$, the corresponding perturbation Hamiltonian reads

\begin{equation}
H_{p}=\sum_{nkqj}\varPsi_{nk+q}^{\dagger}V_{j}^{n}\sigma_{j}\varPsi_{nk}
\end{equation}
where $j=1,2,3,4$ stands for four different densities index in $n$
channel, and $\varPsi_{nk}=\left[\begin{array}{cc}
c_{n1,k}, & c_{n2,-k}^{\dagger}\end{array}\right]^{T}$ is the field operator in Nambu spinor representation. Also with Pauli
matrices $\sigma_{x,y,z}$ and unit matrix $I$ we define $\sigma_{1}\equiv\left(I+\sigma_{z}\right)/2$,
$\sigma_{2}\equiv\left(I-\sigma_{z}\right)/2$, $\sigma_{3}\equiv\left(\sigma_{x}-i\sigma_{y}\right)/2$
and $\sigma_{4}\equiv\left(\sigma_{x}+i\sigma_{y}\right)/2$. This
perturbation Hamiltonian $H_{p}$ introduces a fluctuation of density
matrix 
\begin{equation}
\rho_{q}=\chi V_{{\rm ext}},\label{eq:resp_ks}
\end{equation}
which connects to the external potential $V_{{\rm ext}}$ by the response
function $\chi$ of the system. 
\begin{equation}
\rho_{q}^{n}=\sum_{k}\left[\begin{array}{c}
\varPsi_{nk}^{\dagger}\sigma_{1}\varPsi_{nk+q}\\
\varPsi_{nk}^{\dagger}\sigma_{2}\varPsi_{nk+q}\\
\varPsi_{nk}^{\dagger}\sigma_{3}\varPsi_{nk+q}\\
\varPsi_{nk}^{\dagger}\sigma_{4}\varPsi_{nk+q}
\end{array}\right]
\end{equation}
is the expression of density fluctuation in $n$ channel. Usually
the exact calculation of response function $\chi$ is quite hard and
difficult. However, random phase approximation suggests that we can
treat the fluctuation term of interaction Hamiltonian $H_{{\rm sf}}=\sum_{nq}\rho_{q}^{n\dagger}\cdot A_{q}^{n}$
as a self-consistent dynamical potential $V_{{\rm sf}}^{n}=M_{I}A_{q}^{n},$
in which $A_{q}^{n}=\left[\begin{array}{cccc}
0, & 0, & \Delta_{n,q}, & \Delta_{n,-q}^{*}\end{array}\right]^{T}$is the strength of fluctuation potential with
\begin{equation}
\left[\begin{array}{c}
\Delta_{o,q}\\
\Delta_{o,-q}^{*}\\
\Delta_{c,q}\\
\Delta_{c,-q}^{*}
\end{array}\right]=\sum_{k}\left[\begin{array}{cccc}
V_{oo} & 0 & V_{oc} & 0\\
0 & V_{oo} & 0 & V_{oc}\\
V_{co} & 0 & V_{cc} & 0\\
0 & V_{co} & 0 & V_{cc}
\end{array}\right]\left[\begin{array}{c}
\left\langle \varPsi_{ok}^{\dagger}\sigma_{3}\varPsi_{ok+q}\right\rangle \\
\left\langle \varPsi_{ok}^{\dagger}\sigma_{4}\varPsi_{ok+q}\right\rangle \\
\left\langle \varPsi_{ck}^{\dagger}\sigma_{3}\varPsi_{ck+q}\right\rangle \\
\left\langle \varPsi_{ck}^{\dagger}\sigma_{4}\varPsi_{ck+q}\right\rangle 
\end{array}\right],
\end{equation}
 and $M_{I}$ is just a constant matrix 
\begin{equation}
M_{I}=\left[\begin{array}{cccc}
0 & 0 & 0 & 0\\
0 & 0 & 0 & 0\\
0 & 0 & 0 & 1\\
0 & 0 & 1 & 0
\end{array}\right].
\end{equation}
We should notice that here we neglect the fluctuation from the normal
spin-up and down density. This is reasonable in three dimensional
space where the utilization of $s$-wave contact interaction induces
two anomalous densities to be divergent, and play a much more important
role than normal densities. Now the system can be treated as a mean-field
quasiparticle gases experienced by an effective potential

\begin{equation}
V_{{\rm eff}}\equiv V_{{\rm ext}}+V_{{\rm sf}},\label{eq:veff}
\end{equation}
which induces the density fluctuations $\rho_{q}$ of the system following
\begin{equation}
\rho_{q}=\Pi\cdot V_{{\rm eff}},\label{eq:resp_ks0}
\end{equation}
in which $\Pi$ is the response function matrix in the mean-field
level, whose calculation is feasible and relatively quite easy. Finally,
with Eqs.\ref{eq:resp_ks}, \ref{eq:veff} and \ref{eq:resp_ks0},
we find the desired response function $\chi$ can be calculated by
its connection with mean-field response function $\Pi$ by
\begin{equation}
\chi=\frac{\Pi}{1-\Pi M_{G}},\label{eq:resp_rpa}
\end{equation}
where $M_{G}$ is another constant matrix 
\begin{equation}
M_{G}=\left[\begin{array}{cc}
M_{I}V_{oo}, & M_{I}V_{oc}\\
M_{I}V_{co}, & M_{I}V_{cc}
\end{array}\right].
\end{equation}

The mean-field response function $\Pi$ is a $8\times8$ matrix, its
explicit form is given by

\begin{equation}
\left[\begin{array}{cccccccc}
\Pi_{11}^{o} & \Pi_{12}^{o} & \Pi_{13}^{o} & \Pi_{14}^{o} & 0 & 0 & 0 & 0\\
\Pi_{12}^{o} & \Pi_{22}^{o} & \Pi_{23}^{o} & \Pi_{24}^{o} & 0 & 0 & 0 & 0\\
\Pi_{14}^{o} & \Pi_{24}^{o} & -\Pi_{12}^{o} & \Pi_{34}^{o} & 0 & 0 & 0 & 0\\
\Pi_{13}^{o} & \Pi_{23}^{o} & \Pi_{43}^{o} & -\Pi_{12}^{o} & 0 & 0 & 0 & 0\\
0 & 0 & 0 & 0 & \Pi_{11}^{c} & \Pi_{12}^{c} & \Pi_{13}^{c} & \Pi_{14}^{c}\\
0 & 0 & 0 & 0 & \Pi_{12}^{c} & \Pi_{22}^{c} & \Pi_{23}^{c} & \Pi_{24}^{c}\\
0 & 0 & 0 & 0 & \Pi_{14}^{c} & \Pi_{24}^{c} & -\Pi_{12}^{c} & \Pi_{34}^{c}\\
0 & 0 & 0 & 0 & \Pi_{13}^{c} & \Pi_{23}^{c} & \Pi_{43}^{c} & -\Pi_{12}^{c}
\end{array}\right].\label{eq:resp_mf}
\end{equation}
Different from spin-population balanced case, there are $9$ independent
matrix elements in mean-field response function of each channel, or
$16$ matrix elements in two channels, which make the calculation
of response function become much heavy. In the final appendix of this
paper, we give all expressions of these independent matrix elements.
Although there are no cross terms in matrix of Eq.\ref{eq:resp_mf}
between open and close channel, the coupling effect between these
two channels is indeed existent. First, it is shown in the definition
of two order parameters (Eq.\ref{eq:gap}). Also the random phase
approximation brings a correction to mean-field response function,
which is shown in the denominator of Eq.\ref{eq:resp_rpa}. This correction
brings not only the coupling effect of these two channels, but also
the right correction to investigate all collective excitations. 

Finally, we obtain the intra and inter response functions between
these two channels

\begin{equation}
\begin{array}{c}
\chi^{\left(oo\right)}=\chi_{11}+\chi_{12}+\chi_{21}+\chi_{22},\\
\chi^{\left(oc\right)}=\chi_{15}+\chi_{16}+\chi_{25}+\chi_{26},\\
\chi^{\left(co\right)}=\chi_{51}+\chi_{52}+\chi_{61}+\chi_{62},\\
\chi^{\left(cc\right)}=\chi_{55}+\chi_{56}+\chi_{65}+\chi_{66}.
\end{array}
\end{equation}

\subsection{Dynamic structure factors}

We now turn to discuss dynamic structure factors (DSF for short).
Similarly we can define the intra and inter dynamic structure factor
between two channels, which here is named the total and relative density
dynamic structure factor, respectively \cite{zhang2016}, 

\begin{equation}
S_{{\rm tot}}\left(q,\omega\right)=-\frac{{\rm Im\left[\chi^{\left(oo\right)}+\chi^{\left(oc\right)}+\chi^{\left(co\right)}+\chi^{\left(cc\right)}\right]}}{\pi\left(1-e^{-\hbar\omega/k_{B}T}\right)},
\end{equation}

\begin{equation}
S_{{\rm rel}}\left(q,\omega\right)=-\frac{{\rm Im\left[\chi^{\left(oo\right)}-\chi^{\left(oc\right)}-\chi^{\left(co\right)}+\chi^{\left(cc\right)}\right]}}{\pi\left(1-e^{-\hbar\omega/k_{B}T}\right)},
\end{equation}
where the subscript '{rel}' denotes a relative density oscillation
between the open and close channel. The total density dynamic structure
factor satisfies the famous $f$-sum rule

\begin{equation}
\int d\omega S_{{\rm tot}}\left(q,\omega\right)\omega=N\frac{q^{2}}{2M},
\end{equation}
while at limit $q\rightarrow0$ the relative density dynamic structure
factor satisfies 

\begin{equation}
\int d\omega S_{{\rm rel}}\left(q,\omega\right)\omega=N\frac{q^{2}}{2M}-8\left\langle V_{oc}\right\rangle ,
\end{equation}
where $\langle V_{oc} \rangle$ is the coupling energy between open
and closed channels \cite{zhang2016}. In our following discussion,
we will focus on zero temperature $T=0$, and set Plank constant $\hbar=1$
for simple. 

\section{DSF in balanced BCS superfluid}

Let us first overview a spin-population balanced case with chemical
potential difference $\delta\mu=0$. The ground state of the system
is a BCS-type superfluid in both open and close channel, namely $\left[{\rm BCS}\right]_{o}\left[{\rm BCS}\right]_{c}$.
There are two stable eigenstates, one is in-phase state with order
parameters satisfy $\Delta_{o}\Delta_{c}>0$, and the other one is
an out-of-phase state with $\Delta_{o}\Delta_{c}<0$. Both of them
are possible candidates of ground state, depending on specific interaction
parameters. The collective excitation at transferred momentum $q\ll k_{F}$
or $q\sim k_{F}$ had been discussed in Ref. \cite{zhang2016}. We
will first focus our discussion on a large transferred momentum $q\gg k_{F}$,
and then discuss the situation related to phase transition. 

\begin{figure}
\includegraphics[scale=0.36]{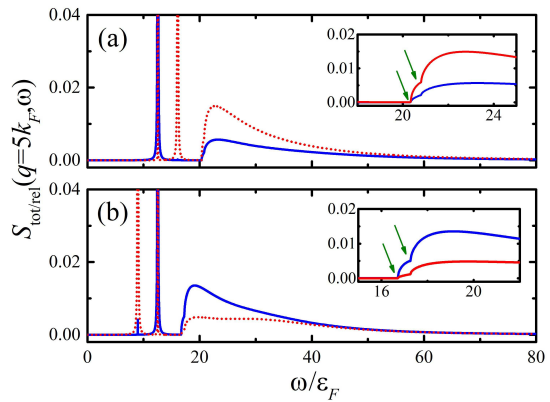}\caption{\label{fig:in_out_dsf} The total density dynamic structure factor
(blue line) and relative one(red line) of BCS superfluid in in-phase
ground state (a) and out-of-phase metastable state (b). Two inset
figures are their own enlarge of the pair-breaking excitation, the
left and right arrow mark the initial position of pair-breaking excitation
in the open and closed channel, respectively.}
\end{figure}

We take interaction parameters $1/k_{F}a_{0}=12/7$ and $1/k_{F}a_{1}=-12$
(or $1/k_{F}a_{-}=2.0$ and $1/k_{F}a_{+}=1.5$), at which the system
is in the BEC regime of the famous BCS-BEC crossover. Then in-phase
state is the ground state, while the out-of-state is an energetical
metastable state. We set Zeeman detuning $\delta\left(B\right)=0.5\varepsilon_{F}$.
The results of total and relative dynamic structure factors of in-phase
and out-of-phase state at transferred momentum $q=5k_{F}$ are shown
in Fig.\ref{fig:in_out_dsf}. Panel (a) is from the in-phase state
with chemical potential $\mu=-3.83\varepsilon_{F}$, open channel
order parameter $\Delta_{o}=1.38\varepsilon_{F}$ and close channel
one $\Delta_{c}=1.23\varepsilon_{F}$, while panel (b) from out-of-phase
state with $\mu=-2.05\varepsilon_{F}$, $\Delta_{o}=1.03\varepsilon_{F}$
and $\Delta_{c}=-1.23\varepsilon_{F}$. Something in common in these
two panels is that, all dynamic structure factors consist of four
different excitations, which are molecular excitation, collective
Leggett mode and Cooper-pair breaking excitation in both open and
close channel. The molecular excitation of both panels always locates
at $q^{2}/4M\simeq12.5\varepsilon_{F}$, which are only dependent
on the value of transferred momentum $q$, no matter what state it
is. Since the chemical potential in two states both satisfy $\mu<0$,
the minimum energy required to break a Cooper-pair is $\omega_{th}^{\left(o\right)}=2\sqrt{(q^{2}/8M-\mu)^{2}+\Delta_{o}^{2}}$
in open channel and $\omega_{th}^{\left(c\right)}=2\sqrt{(q^{2}/8M-\mu+\delta\left(B\right)/2)^{2}+\Delta_{c}^{2}}$
in close channel, respectively. These two initial excitations generate
two twist points in two inset figures, located by two arrows. Panel
(a) is the ground state because the in-phase state requires more energy
to break a Cooper-pair than the case of out-of-phase state shown in
Panel (b). This is also proved by a relatively longer energy distance
between molecular excitation and ${\rm min}\left[\omega_{th}^{\left(o\right)},\omega_{th}^{\left(c\right)}\right]$,
whose value is almost equal to the bigger bound energy $\varepsilon_{b}^{-}=-\hbar^{2}/Ma_{-}^{2}=8.0\varepsilon_{F}$,
while the energy distance between Leggett mode and ${\rm min}\left[\omega_{th}^{\left(o\right)},\omega_{th}^{\left(c\right)}\right]$
take the other smaller bound energy $\varepsilon_{b}^{+}=-\hbar^{2}/Ma_{+}^{2}$
(these two energy differences will be exactly equal to $\varepsilon_{b}^{\pm}=-\hbar^{2}/Ma_{\pm}^{2}=4.5\varepsilon_{F}$
when $\delta\left(B\right)=0$). This is also the reason why the Leggett
mode locates on the right hand side of molecular excitation in Panel
(a), or on the left hand side in Panel (b). Here dynamic structure
factors display rich information about large transferred momentum
dynamical excitations, which do a great help to distinguish ground
in-phase state from metastable out-of-phase state.

\begin{figure}
\includegraphics[scale=0.36]{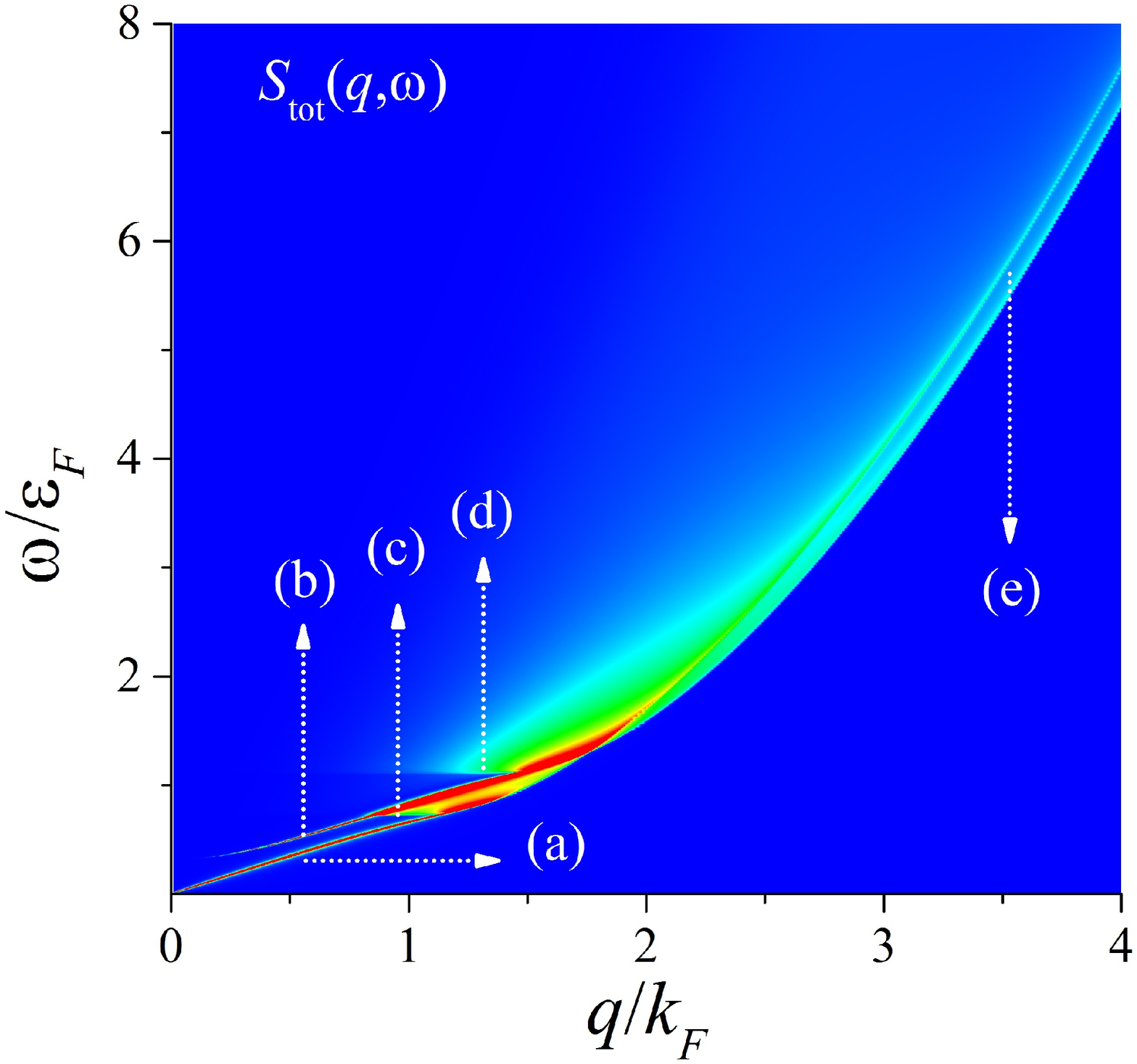}

\includegraphics[scale=0.36]{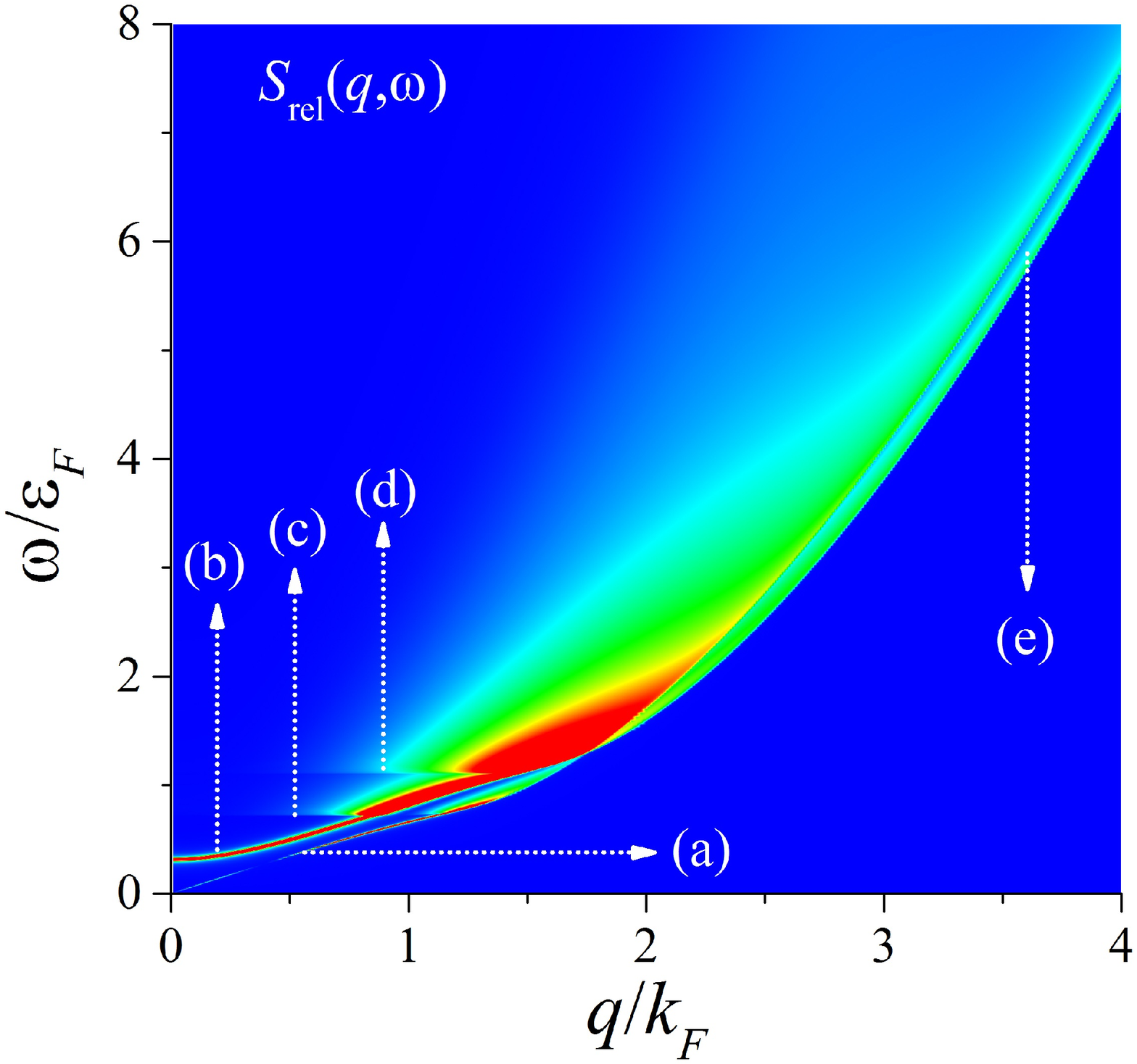}

\caption{\label{fig:BCS_dsf} The total (upper panel) and relative (lower panel)
dynamic structure factor in BCS superfluid side of phase transition,
with scattering length $\ensuremath{1/k_{F}a_{0}=-0.5}$, $\ensuremath{1/k_{F}a_{1}=-0.05}$,
Zeeman detuning $\delta(B)=0.4\varepsilon_{F}$. The result is independent
on the value of chemical potential difference when $\delta\mu<\delta\mu_{{\rm cr}}$.
(a) phonon mode, (b) Leggett mode, (c) $2\Delta_{c}$, (d) $2\Delta_{o}$,
(e) pair-breaking excitations at a large $q$ in two channels.}

\end{figure}

Next we discuss the phase transition from BCS to Sarma superfluid
with the same parameters in Fig.\ref{Fig1_phase}. When chemical potential
difference $\delta\mu$ is smaller than the critical value $\delta\mu_{{\rm cr}}=0.486\varepsilon_{F}$,
a BCS superfluid state becomes the ground state of the system. These
parameters are like the results of single-channel Fermi superfluid
close to unitary regime of BCS-BEC crossover, with a positive chemical
$\mu=0.43\varepsilon_{F}$. As shown in Fig.\ref{fig:BCS_dsf}, at
a small transferred momentum $q$, we observe two collective excitations
(red curves in two panels), the phonon (a) and Leggett mode (b). Relatively,
the gapless phonon mode is easier to investigate in total dynamic
structure factor than in the relative dynamic structure factor, while
the gapped Leggett mode just in inverse. Two short horizontal lines
at $\omega^{\left(c\right)}=2|\Delta_{c}|=0.72\varepsilon_{F}$ (c)
and $\omega^{\left(o\right)}=2|\Delta_{o}|=1.10\varepsilon_{F}$ (d)
represent two minimum values of energy to break a Cooper pair in open
and close channel, respectively, while the same physics evolves into
a two-peak signal at a large $q$ (e). Here in fact all results are
independent the value of chemical potential difference $\delta\mu$,
when $\delta\mu<\delta\mu_{{\rm cr}}$. The system is always a spin-population
balanced BCS superfluid state.

\section{DSF in Sarma superfluid}

\begin{figure}
\includegraphics[scale=0.36]{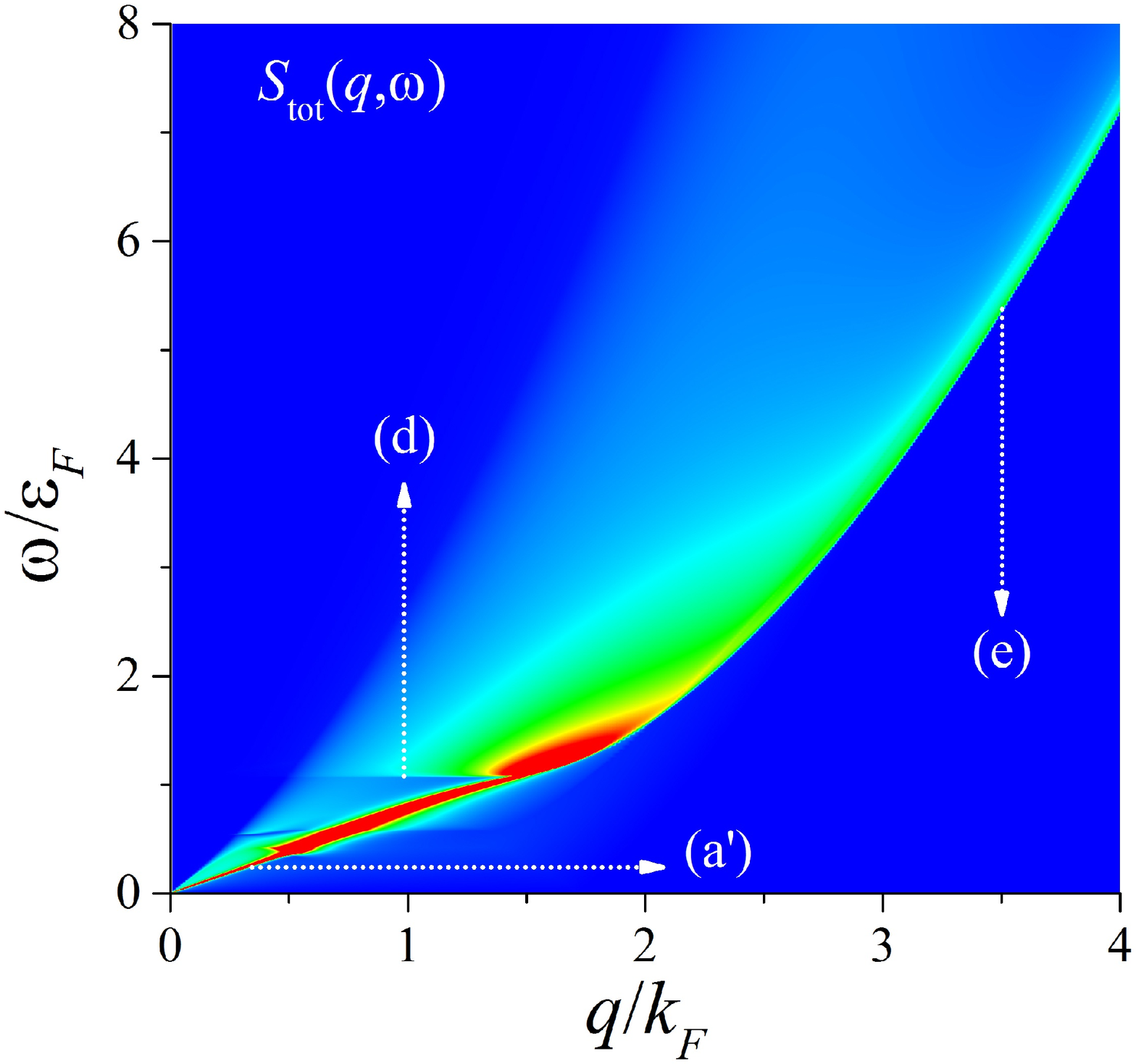}

\includegraphics[scale=0.36]{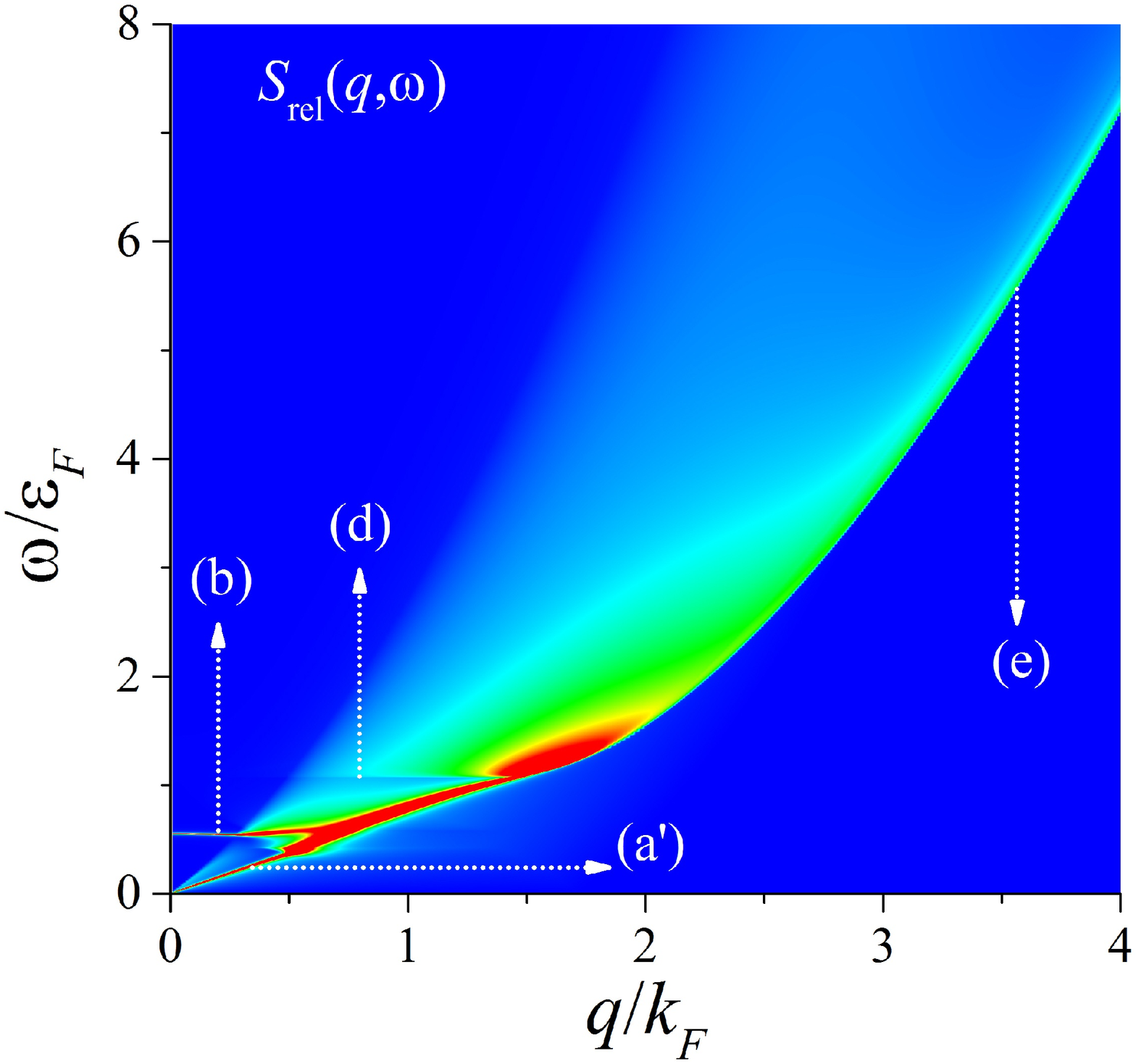}

\caption{\label{fig:Sarma_dsf} The total (upper panel) and relative (lower
panel) dynamic structure factor in Sarma superfluid side of phase
transition, with scattering length $\ensuremath{1/k_{F}a_{0}=-0.5}$,
$\ensuremath{1/k_{F}a_{1}=-0.05}$, Zeeman detuning $\delta(B)=0.4\varepsilon_{F}$,
and chemical potential difference $\delta\mu=0.5\varepsilon_{F}$.
(${\rm a'}$) phonon mode mixed with single-particle gapless excitation
in close channel, (b) Leggett mode, (d) $2\Delta_{o}$, (e) pair-breaking
excitations at a large $q$ in two channels.}
\end{figure}

\begin{figure}
\includegraphics[scale=0.36]{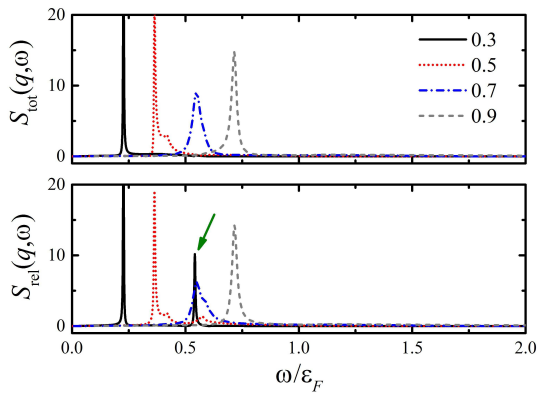}\caption{\label{fig:sarma_qs} The total (upper panel) and relative (lower
panel) dynamic structure factors of Sarma superfluid at small transferred
momentums, $q=0.3,0.5,0.7,0.9k_{F}$. A green arrow in lower panel
locates the Leggett mode.}
\end{figure}

\begin{figure}
\includegraphics[scale=0.36]{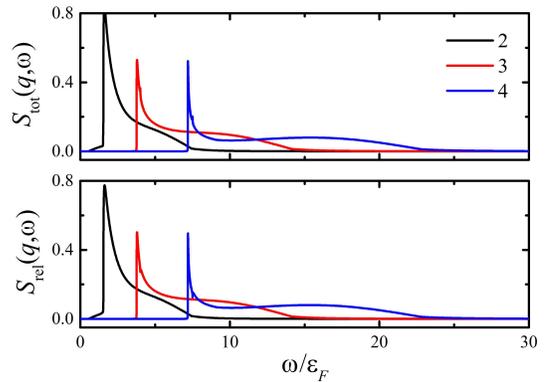}\caption{\label{fig:sarma_qb} The total (upper panel) and relative (lower
panel) dynamic structure factors of Sarma superfluid at big transferred
momentums, $q=2,3,4k_{F}$.}
\end{figure}

For the same scattering length $\ensuremath{1/k_{F}a_{0}=-0.5}$,
$\ensuremath{1/k_{F}a_{1}=-0.05}$ and Zeeman detuning $\delta(B)=0.4\varepsilon_{F}$,
but varying the chemical potential difference over the critical value
$\delta\mu_{{\rm cr}}$ to $\delta\mu=0.5\varepsilon_{F}$, the system
comes into a spin-population polarized Sarma superfluid state $\left[{\rm BCS}\right]_{o}\left[{\rm Sarma}\right]_{c}$.
The chemical potential difference $\delta\mu$ is larger than the
order parameter in close channel $\Delta_{c}=-0.08\varepsilon_{F}$,
but smaller than one in open channel $\Delta_{o}=0.54\varepsilon_{F}$.
Then the close channel becomes a gapless Sarma superfluid, while the
open channel is a BCS superfluid. Our main results of the total and
relative dynamic structure factor in this Sarma superfluid are shown
respectively in two panels of Fig.\ref{fig:Sarma_dsf}, in the form
of two-dimensional contour plot in the $q$-$\omega$ plane. The dynamic
structure factors as a function of $\omega$ at small and large transferred
momenta are reported in Fig.\ref{fig:sarma_qs} and Fig.\ref{fig:sarma_qb},
respectively. 

Different from the results of BCS superfluid whose phonon excitation
has an obvious delta-like structure, the phonon excitation here has
a finite expansion width because of the scattering with the gapless
single-particle excitation in the close channel, which is denoted
as (${\rm a'}$) in Fig.\ref{fig:Sarma_dsf}. In greater detail, all
collective excitations are focused in a small transferred momentum
$q$ regime $\left(q<k_{F}\right)$. A single-particle gapless excitation
of Sarma superfluid can be observed in both total and relative dynamic
structure factors, and it is mixed with another gapless excitation,
the collective phonon excitation. This monotonically increasing linear
behavior of dynamic structure factor is also displayed in Fig.\ref{fig:sarma_qb}
at $q=2k_{F}$, where the transferred energy starts from a very small
value. The results in this regime are quite similar to behavior of
ideal Fermi gases \cite{mazzanti1996}, and it can be understood from
the fact that some fermionic atoms are not paired, and behave like
free fermions in close channel. This gapless dynamical behavior is
the most specific character of Sarma superfluid. 

Also we can observe an horizontal gapped Leggett excitation, whose
position almost does not vary with the transferred momentum $q$.
This is also shown in the peak around $\omega\simeq0.6\varepsilon_{F}$
of relative dynamic structure factor in Fig.\ref{fig:sarma_qs}, located
by a green arrow. Going on increasing transferred momentum $q$, a
merge procedure between Leggett and phonon mode can be clearly investigated
in the relative dynamic structure factor (Panel (b) of Fig.\ref{fig:Sarma_dsf},
and lower panel of Fig.\ref{fig:sarma_qs}), the critical value of
transferred momentum is around $q\approx0.7k_{F}$, after which the
phonon and Leggett excitations are mixed together and show only a
wide peak, as shown by gray dot line in Fig.\ref{fig:sarma_qs}. 

Another physics we are interested in is the minimum value to break
Cooper pairs in two channels. Different from the BCS superfluid, only
one short horizontal line at $\omega^{(o)}=2\Delta_{o}=1.08\varepsilon_{F}$,
who is the minimum energy to break a Cooper-pair in open channel,
is observed, while the other one in close channel is completely immersed
in the gapless excitation.

\begin{figure}
\includegraphics[scale=0.36]{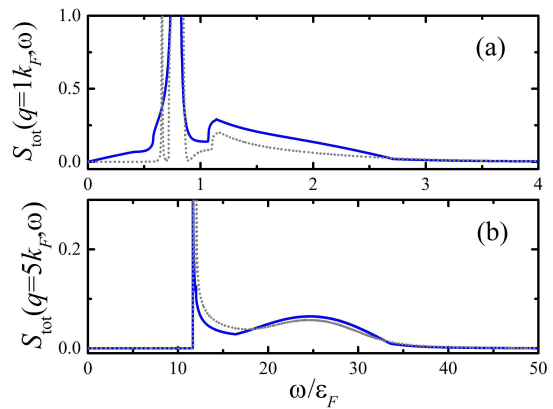}\caption{\label{fig:tot_q15} The comparison of total dynamic structure factor
between BCS superfluid state (blue line) and Sarma superfluid state
(grey dot line) at $q=1k_{F}$ (a) and $q=5k_{F}$ (b).}
\end{figure}

At a large $q$, the molecular excitation and atomic excitation are
both observed in Fig.\ref{fig:sarma_qb} and Panel (b) of Fig.\ref{fig:tot_q15}.
Different from the BCS superfluid, the molecular excitation does not
have a clear two-peak signal related to pair-breaking excitations
in two channels, which is because that the Cooper pair breaking in
close channel is quite weak, compared with one in open channel. We
can only observe a very small peak in total dynamic structure factors,
or a very small dip in relative dynamic structure factors, whose location
is after the first big Cooper-pair breaking peak in Fig.\ref{fig:sarma_qb}.
The atomic excitation, whose location is around $q^{2}/2M$, becomes
more and more obvious when increasing transferred momentum $q$. Also
Sarma superfluid has a relative stronger atomic excitation than one
in BCS superfluid since it has more surplus of unpaired Fermi atoms.

Finally we give Fig.\ref{fig:tot_q15} by which to present the main
difference between BCS superfluid (gray line) and Sarma superfluid
(blue line). Compared with BCS superfluid, Sarma superfluid has a
linear gapless excitation at $q=k_{F}$, mixed with the collective
phonon excitation (Panel (a)), and a stronger atomic excitation (Panel
(b)) at $q=5k_{F}$.

\section{Conclusions}

In summary, we investigate dynamic structure factors of a two-channel
Fermi superfluid with orbital Feshbach resonance. In spin-population
balanced BCS superfluid state, dynamic structure factors can do help
to distinguish the ground in-phase from metastable out-of-phase state
by the location of minimum energy to break a Cooper-pair, or the relative
location between collective Leggett mode and molecular excitation
in a BEC-like parameters regime. For another different interaction
parameters where Sarma superfluid may replace the BCS superfluid as
a possible ground state of the system, we numerically calculate dynamic
structure factors and find their dynamical characters in both spin-population
balanced BCS superfluid and spin-polarized Sarma superfluid. An obvious
gapless excitation at a small transferred energy is observed only
in Sarma superfluid, which also has a stronger atomic excitation at
a large transferred momentum. These discoveries can serve to differentiate
all possible eigenstates of the system. 

\section{Acknowledgement}

Our research was supported by the National Natural Science Foundation
of China, Grants No. 11804177 (P.Z.), No. 11547034 (H.Z.) and No.
11775123 (L.H.), by the Shandong Provincial Natural Science Foundation,
China, Grant No. ZR2018BA032 (P.Z.), and by Australian Research Council\textquoteright s
(ARC) Discovery Projects No. DP180102018 (X.-J.L), and No. DP170104008
(H.H.). L.H. acknowledges the support of the Recruitment Program for
Young Professionals in China (i.e., the Thousand Young Talent Program).

\subsection{Appendix}

In this appendix, we will give the derivation of mean-field response
functions $\Pi$ in both open and close channels. Actually the response
functions in these two channels are the same as each other, but with
their corresponding values of chemical potential $\mu_{n}$ and order
parameter $\Delta_{n}$ , where $n=o,c$ is the channel index. For
example, in $n$ channel, there are $9$ independent matrix elements,
which are expressed as 

\[
\Pi_{11}^{n}\left(q,\omega\right)=+\sum_{kll'}P_{1}^{l}\left(k\right)P_{1}^{l'}\left(k+q\right)F_{k}^{ll'}\left(q,\omega\right),
\]

\[
\Pi_{12}^{n}\left(q,\omega\right)=-\sum_{kll'}P_{3}^{l}\left(k\right)P_{3}^{l'}\left(k+q\right)F_{k}^{ll'}\left(q,\omega\right),
\]

\[
\Pi_{13}^{n}\left(q,\omega\right)=-\sum_{kll'}P_{1}^{l}\left(k\right)P_{3}^{l'}\left(k+q\right)F_{k}^{ll'}\left(q,\omega\right),
\]

\[
\Pi_{14}^{n}\left(q,\omega\right)=-\sum_{kll'}P_{3}^{l}\left(k\right)P_{1}^{l'}\left(k+q\right)F_{k}^{ll'}\left(q,\omega\right),
\]

\[
\Pi_{22}^{n}\left(q,\omega\right)=+\sum_{kll'}P_{2}^{l}\left(k\right)P_{2}^{l'}\left(k+q\right)F_{k}^{ll'}\left(q,\omega\right),
\]

\[
\Pi_{23}^{n}\left(q,\omega\right)=+\sum_{kll'}P_{3}^{l}\left(k\right)P_{2}^{l'}\left(k+q\right)F_{k}^{ll'}\left(q,\omega\right),
\]

\[
\Pi_{24}^{n}\left(q,\omega\right)=+\sum_{kll'}P_{2}^{l}\left(k\right)P_{3}^{l'}\left(k+q\right)F_{k}^{ll'}\left(q,\omega\right),
\]

\[
\Pi_{34}^{n}\left(q,\omega\right)=+\sum_{kll'}P_{2}^{l}\left(k\right)P_{1}^{l'}\left(k+q\right)F_{k}^{ll'}\left(q,\omega\right),
\]

\[
\Pi_{43}^{n}\left(q,\omega\right)=+\sum_{kll'}P_{1}^{l}\left(k\right)P_{2}^{l'}\left(k+q\right)F_{k}^{ll'}\left(q,\omega\right),
\]
where 

\[
P_{1}^{l}\left(k\right)=\frac{1}{2}\left(1+l\frac{\xi_{nk}}{E_{nk}}\right),
\]

\[
P_{2}^{l}\left(k\right)=\frac{1}{2}\left(1-l\frac{\xi_{nk}}{E_{nk}}\right),
\]

\[
P_{3}^{l}\left(k\right)=l\frac{\Delta_{n}}{2E_{nk}},
\]

\[
F_{k}^{ll'}\left(q,\omega\right)=\frac{f\left(E_{nk}^{l}\right)-f\left(E_{nk+q}^{l'}\right)}{\omega+i\eta+E_{nk}^{l}-E_{nk+q}^{l'}}.
\]
$q$ and $\omega$ are respectively the transferred momentum and energy,
$l=\pm1$ stands for the particle or hole solution, $\eta$ is a small
positive number. Specifically one should notice that expressions of
$\Pi_{34}^{n}$ and $\Pi_{43}^{n}$ are both divergent. The random
phase approximation, shown in the denominator of Eq.\ref{eq:resp_rpa},
do help to cure the divergence of them.

\end{document}